\documentclass[11pt]{rmta2013esp}
\paginas{1}{9}
\newtheorem{definition}{Definici\'on}
\newtheorem{theorem}{Teorema}
\newtheorem{lemma}{Lema}

\newtheorem{corollary}{Corolario}

\usepackage{amsmath}
\usepackage{amsfonts}
\usepackage{latexsym}
\usepackage{graphicx}
\usepackage{url}
\usepackage[english]{babel}

\newcommand{\HRule}{\begin{center}\rule{5cm}{0.2mm}\end{center}}

\begin{document}

\title{INTERSTATIS: The STATIS method for interval valued data}

\author{Oldemar Rodr\'{\i}guez \thanks{University of Costa Rica, San Jos\'e, Costa Rica;
E-Mail: oldemar.rodriguez@ucr.ac.cr}
-- David Corrales\thanks{David Corrales, HP Networking R\&D; E-Mail:  david.corrales@hp.com}}
\date{Recibido: 27 de mayo 2013}

\maketitle

\begin{resumen}
El m\'etodo STATIS, propuesto por L'Hermier des Plantes
y Escoufier, se utiliza para analizar m\'ultiples tablas de datos en las cuales
es muy frecuente que cada una de la tablas tenga informaci\'on referente al mismo conjunto de individuos. Las diferencias y similitudes
entre dichas tablas se analizan por medio de una estructura llamada {\em compromiso}.
En este trabajo se presenta un nuevo algoritmo, denominado INTERSTATIS, del m\'etodo STATIS
para el caso cuando los datos de entrada son todos de tipo intervalo. Esta propuesta
se basa en la aritm\'etica de intervalos de Moore y el M\'etodo de Centros
para el An\'alisis de Componentes Principales con datos de tipo intervalo, propuesto por Cazes et al. \cite{cazes1997}.
Adem\'as de presentar el m\'etodo INTERSTATIS de forma algor\'{\i}tmica,
un ejemplo de ejecuci\'on es presentado, junto con la interpretaci\'on de sus resultados.
\end{resumen}

\PC
INTERSTATIS, STATIS, Aritm\'etica de intervalos, datos tipo intervalo, An\'alisis de datos simb\'olico.

\begin{abstract}
The STATIS method, proposed by L'Hermier des Plantes
and Escoufier, is used to analyze multiple data tables in which
is very common that each of the tables have information concerning the same set of individuals.
The differences and similitudes
between said tables are analyzed by means of a structure called the
\emph{compromise}.
In this paper we present a new algorithm for applying the STATIS
method when the input consists of interval data. This proposal
is based on Moore's interval arithmetic and the Centers Method
for Principal Component Analysis with interval data, proposed by Cazes el al. \cite{cazes1997}.
In addition to presenting the INTERSTATIS method in an algorithmic way,
an execution example is shown, alongside the interpretation of its results.
\end{abstract}

\KW
INTERSTATIS, STATIS, Interval Arithmetic, Interval Data, Symbolic data analysis.
\medskip


\section{Introduction}
Given the flexibility and various applications of the Principal Component Analysis (PCA),
L'Hermier des Plantes \cite{lhermier1976} and Escoufier \cite{escoufier1980, lavit1994}
worked, sequentially, in a generalized version of the method, to study multiple data tables.
The method, named \texttt{STATIS}, is capable of analyzing several groups of tables,
referring to the same variables or the same individuals (\texttt{STATIS DUAL}).
The \texttt{STATIS} uses the PCA as part of the necessary transformations to reach a
structure called the \emph{compromise}, used in evaluating the differences and similitudes
amongst the input tables.

Taking into account the potential of the \texttt{STATIS}, as an 3-index information analysis tool,
the initiative of extending the method to the case of interval symbolic objects \cite{Billard2011, Makosso2012, billard2006, diday1987}
was born. These objects are meant to represent second order inviduals, whose given \emph{concepts}
are much more complex than simple points in $\mathbb{R}^n$. For example, interval vectors,
histograms, graphs, trees, rules, sets, functions, etc.

Although there already exist works in the field of symbolic methods for data analysis,
like the generalizations of the Principal Components Analysis (PCA) \cite{cazes1997, lauro2000, rodriguez2001} and the
$k$-means \cite{verde2000} among others, the \texttt{STATIS} method has not
been reformulated to handle symbolic objects. The first step consists in finding a way
to generalize the \texttt{STATIS}, while guaranteeing that the classic method is an specific case of
the symbolic proposal. This restriction is important since it ensures that the proposal
has a sound statistical meaning.

Initially, Moore's Interval Arithmetic (IA) \cite{caprani2002, moore1979} was proposed to
perform operations between intervals. This arithmetic was created to handle, in a
bounded way, the uncertainty in operations amongst non-exact quantities, and thus,
provides a natural transition for the elemental operations --addition, substraction,
division and multiplication-- from the classic case to the symbolic one. Besides,
it is simple in its definition, implementation and complies with the fact that the
operations between real numbers are specific cases of the IA, making it a good
starting point for the generalization of the \texttt{STATIS}.

Additionally, knowing that the PCA is an integral part of the \texttt{STATIS},
it was necessary to have a PCA variant capable of handling interval valued data.
To solve this need the Centers PCA (CPCA) by Cazes et al. \cite{cazes1997} was used.
This PCA gives as a result a representative hypercube for each individual, where its
variation (in width, depth and height) describes the correspondent individual's variation.

Besides the regular interval notation, the $\widehat\times$ notation is used to indicate
interval arithmethic operations. In the case of the dot product between symbolic matrices,
its definition is analog to the dot product between real matrices, but using the IA multiplication.

Note as well the use of boldface to denote interval valued variables. For example,
$\mathbf{X}$ denotes an interval matrix, while $X$ stands for a classic matrix.
This provision facilitates distinguishing between both cases throughout the paper.

\section{INTERSTATIS with IA and Centers PCA}
This first version of the \texttt{INTERSTATIS} for interval data, works by replacing
the conventional --or classic-- operations, with their IA counterparts. Also, the CPCA,
with its interval inputs and outputs, replaces the classic PCA method.

During the preprocessing phase of the input matrices, it is a requisite to center them.
For this step the Rodr\'iguez \cite{rodriguez2000b} intervals mean was used. It follows:
\begin{definition}[Intervals Mean]
\label{promedioIntervalos}
Let $\mathbf{Y}$ be an interval valued variable, defined over $E=\{1,2,\ldots,n\}$
for $\mathbf{Y} = \{[\underline{y}_1, \overline{y}_1]\, [\underline{y}_2, \overline{y}_2], \ldots, [\underline{y}_n, \overline{y}_n]\}$,
the mean is defined as:
\[
\overline{\mathbf{Y}} = \left[ \frac{1}{n} \sum_{i=1}^{n}\underline{y}_i , \frac{1}{n} \sum_{i=1}^{n}\overline{y}_i\right]\text{.}
\]
\end{definition}

Following, the \texttt{INTERSTATIS} algorithm using IA and CPCA, as a generalization
of the version presented by Gonz\'alez and Rodr\'iguez in \cite{gonzalez1995}.\\

\HRule
\begin{center}
\noindent \textbf{INTERSTATIS with AI and CPCA}
\end{center}

\noindent\textbf{Input}: $r$ data tables (matrices)
$\mathbf{X}_k \in \mathcal{I}(\mathbb{M}^{n \times p_k})$ individuals$\times$variables
of size $n \times p_k$, centered with respect to $D$ (using
Definition \ref{promedioIntervalos} for the means), where
\begin{itemize}
\item $n$ is the number of individuals, the same in the $r$ measurings.

\item $p_k$ is the number of variables in the $k$th measuring.

\item $D = diag(1/n, 1/n, \ldots, 1/n)$ is the weights matrix, invariant for the
$r$ measurings.

\item $l = \sum_{k=1}^{r} p_k$ is the amount of variables in all tables.
\end{itemize}

\noindent\textbf{Output}: the following tables are generated:
\begin{itemize}
\item $ \mathbf{T} \in \mathcal{I}(\mathbb{M}^{r \times r})$ with the correlations between tables (step 4).

\item $\mathbf{E_v} \in \mathcal{I}(\mathbb{M}^{l \times l})$ with the evolution of the variables (step 7).

\item $\mathbf{M_i} \in \mathcal{I}(\mathbb{M}^{n \times l})$ with the average individuals (step 7).

\item $\mathbf{E_i} \in \mathcal{I}(\mathbb{M}^{rn \times l})$ with the evolution of the individuals (step 11).
\end{itemize}

\noindent\textbf{Interstructure}: (steps 1-4). Given that the individuals in the $r$ measurings are the same,
it is possible to compare their spatial distribution by means of the $\mathbf{W}_i = \mathbf{X}_i \widehat \times \mathbf{X}_i^T$
matrices. For this, the metric defined by the internal product is used
\begin{displaymath}
<\mathbf{W}_i, \mathbf{W}_j>_\phi = trace(\mathbf{W}_i \widehat\times \mathbf{W}_j) = <\vec{\mathbf{W}_i} \widehat \times \vec{\mathbf{W}_j}>_I
\end{displaymath}
where $\vec{\mathbf{W}_i} \in \mathcal{I}(\mathbb{R}^{n^2})$ is the vector formed
by all the rows of the $\mathbf{W}_i$ matrix, for $i = 1, 2, \ldots, r$.

\begin{list}{}{}
\item \textbf{Step 1}: Compute the $\mathbf{W}_i \in \mathcal{I}(\mathbb{M}^{n \times n})$ matrices,
given by $\mathbf{W}_i = \mathbf{X}_i \widehat \times \mathbf{X}_i^T$ for $i = 1, 2, \ldots, r$.

\item \textbf{Step 2}: Generate the following matrix
\begin{displaymath}
\mathbf{X} = \left[\ \vec{\mathbf{W}_1} \ |\ \vec{\mathbf{W}_2} \ |\ \cdots |\ \vec{\mathbf{W}_r} \ \right]_{\mathcal{I}(n^2 \times r)}
\end{displaymath}

\item \textbf{Step 3}: Perform a CPCA for the triplet $(\mathbf{X}, D_{\frac{1}{\sigma^2}}, \frac{1}{n^2} I_n^2)$.

Store the resulting principal components as
\begin{displaymath}
\mathbf{T} =\left[\mathbf{PC}_1, \mathbf{PC}_2, \ldots, \mathbf{PC}_r\right]_{\mathcal{I}(r \times r)}\text{.}
\end{displaymath}

Also store the first eigenvector, noted by $u = (u_1, u_2, \ldots, u_r)$, and the first eigenvalue $\lambda_1$,
both real, to use them later on in the creation of the \emph{compromise}.
It is not necessary to convert these to interval data since the IA defines operations between
interval and real values.

\item \textbf{Step 4}: Plot the correlations circle from the $\mathbf{T}$ matrix.
Each point represents a data table, where close points mean similar individual configurations,
and parallel vectors represent homothetic configurations.
\end{list}

\noindent\textbf{Intrastructure}: (steps 5-12). The evolutive study of the individuals
and the variables is performed through the \emph{compromise} $\sum_{k=1}^{r} \mathbf{\beta}_k \mathbf{W}_k$
product of the CPCA with parameters $(\mathbf{\tilde X}, I_l, D)$, where $\mathbf{\beta}_k$ is defined in
step 5 and $\mathbf{\tilde X}$ is defined as:
\begin{displaymath}
\mathbf{\tilde X} = \left[\ \mathbf{\beta}_1 \widehat{\times} \mathbf{X}_1 \ |\ \mathbf{\beta}_2 \widehat{\times} \mathbf{X}_2 \ |\ \cdots |\ \mathbf{\beta}_r \widehat{\times} \mathbf{X}_l \right]_{\mathcal{I}(n \times l)}
\end{displaymath}

\begin{list}{}{}
\item \textbf{Evolution of the variables}: (steps 5-9)
\begin{list}{}{}
\item \textbf{Step 5}: Compute $\mathbf{\beta} = (\mathbf{\beta}_1, \mathbf{\beta}_2, \ldots, \mathbf{\beta}_r) = \frac{1}{\sqrt{\lambda_1}}u$.

\item \textbf{Step 6}: Compute the blocks matrix $\mathbf{\tilde X}$ as follows:
\begin{displaymath}
\mathbf{\tilde X} = \left[\ \mathbf{\beta}_1 \widehat{\times} \mathbf{X}_1 \ |\ \mathbf{\beta}_2 \widehat{\times} \mathbf{X}_2 \ |\ \cdots |\ \mathbf{\beta}_r \widehat{\times} \mathbf{X}_l \right]_{\mathcal{I}(n \times l)}
\end{displaymath}

\item \textbf{Step 7}: Perform a CPCA for ($\mathbf{\tilde X}, I_l, D$).
Store the principal components $\mathbf{M_i}$ and the principal correlations
$\mathbf{E_v}$, in the following way
\begin{displaymath}
\mathbf{M_i} = \left[\mathbf{C}_1, \mathbf{C}_2, \ldots, \mathbf{C}_l\right]_{\mathcal{I}(n \times l)} \text{\quad  y\quad}
\mathbf{E_v} = \left[\mathbf{PC}_1, \mathbf{PC}_2, \ldots, \mathbf{PC}_n\right]_{\mathcal{I}(l \times l)} \text{.}
\end{displaymath}

\item \textbf{Step 8}: Plot the correlations circle from $\mathbf{E_v}$,
in which it is possible to study the evolution of the variables.

\item \textbf{Step 9}: Plot the principal plane based on $\mathbf{M_i}$,
in which the $n$ average individuals (see definition
in \cite{lavit1988}) are represented.
\end{list}

\item \textbf{Evolution of the individuals}: (steps 10-12)
\begin{list}{}{}
\item \textbf{Step 10}: Compute the \textbf{\texttt{IND}} matrix, defined by blocks as:
\begin{displaymath}
\texttt{IND} =
\left[
\begin{array}{c}
\mathbf{W}_1 \\
\mathbf{W}_2 \\
\vdots \\
\mathbf{W}_r \\
\end{array}
\right]_{\mathcal{I}(rn \times n)}
\end{displaymath}

\item \textbf{Step 11}: Compute the coordinates for the individuals throughout the
different tables, noted as $\mathbf{E_i}$, through the following matrix product:
\begin{displaymath}
\mathbf{E_i} = \texttt{IND} \widehat \times \mathbf{M_i}
\end{displaymath}

\item \textbf{Step 12}: Plot the principal plane using $\mathbf{E_i}$.
In this chart, one can study the evolution of the individuals.
\end{list}
\end{list}

\HRule

The following results prove that the classical \texttt{STATIS} is a particular
case of the \texttt{INTERSTATIS} with IA and CPCA.

\begin{definition}[Classical and interval matrix equivalence]
\label{equivalenciaMatrices}
Let $X \in \mathbb{M}^{n \times p}$ and $\mathbf{Y} \in \mathcal{I}(\mathbb{M}^{n \times p})$.
Then, $X$ is matrix-equivalent to $\mathbf{Y}$, noted by $X \equiv \mathbf{Y}$ if
\begin{displaymath}
X \equiv \mathbf{Y} \iff \forall\ x_{ij} \in X, \forall\ \mathbf{y}_{ij} \in \mathbf{Y}: x_{ij} = \underline{\mathbf{y}_{ij}} = \overline{\mathbf{y}_{ij}}
\end{displaymath}
with $i=\{1,2,\ldots,n\}$ and $j=\{1,2,\ldots,p\}$.
\end{definition}

\begin{lemma}
\label{casoParticularACPCentros}
The classic PCA is a particular case of the CPCA \cite{rodriguez2000b}.
\end{lemma}

\begin{lemma}
\label{casoParticularAI}
The classic operations $\{+,-,\times, /, \sqrt{\cdot}\}$ are
particular cases of the IA \cite{moore1979}.
\end{lemma}

\begin{theorem}
\label{casoParticularINTERSTATISAI}
The classic $\emph{\texttt{STATIS}}$ is a particular case of the $\emph{\texttt{INTERSTATIS}}$ with IA and CPCA.
\end{theorem}
Due to space concerns the proof has been ommitted, although it is trivial.

\begin{corollary}
\label{resultadoStatis}
Let $X \in \mathbb{M}^{n \times p}$ and $\mathbf{Y} \in \mathcal{I}(\mathbb{M}^{n \times p})$ with
$X \equiv \mathbf{Y}$, then the output of the $\emph{\texttt{STATIS}}$ taking $X$ as input,
is the same as the $\emph{\texttt{INTERSTATIS}}$ output, taking $\mathbf{Y}$ as input.
\end{corollary}

\section{Experiments}
\label{ch:experiments}
This section is concerned with the data used in the tests and the interpretation
of the correspondent results.

For the experiments a data table based on the wine data by Abdi \cite{abdi2006}
was generated. This table describes the criteria from three different
wine experts, on six different types of wines, depending on the
type of oak barrels used in their aging process. For each wine sample,
the intensity of a given flavor is defined in a scale from 1 to 10.
Also, for the generated table, the center of each one of its entries
is the same as the correspondent entry in the original data table.
Mathematically, let $X$ be the classic wines table and $\mathbf{Y}$
the test data matrix, then
\begin{displaymath}
\forall\ x_{ij} \in X, \forall\ \mathbf{y}_{ij} \in \mathbf{Y} : x_{ij} = m(\mathbf{y}_{ij}),
\end{displaymath}
with $i=\{1,2,\ldots,n\}$, $j=\{1,2,\ldots,p\}$ and where $m(\mathbf{y})$
is the middle point or center of the interval $\mathbf{y}$.

This property for the test table is based on the way the CPCA works.
This method projects, for each interval, the
center and then the \emph{minima} and \emph{maxima} in
supplementary. Therefore, if the center --for each entry-- of the
test table is equal to the value in the original table, the individuals
will have a similar spatial placement in the experiment.

\subsection{Interpretation of the results}
Each of the figures presents the output of the \texttt{INTERSTATIS}
using the test table, with the first two axes used for plotting. The charts in this paper were generated in Mathematica.

\begin{figure}[htbp]
\begin{center}
\begin{tabular}{cc}
\includegraphics[width=3cm]{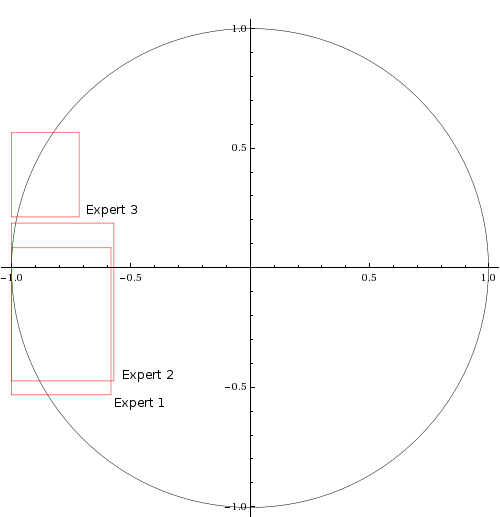} & \includegraphics[width=3cm]{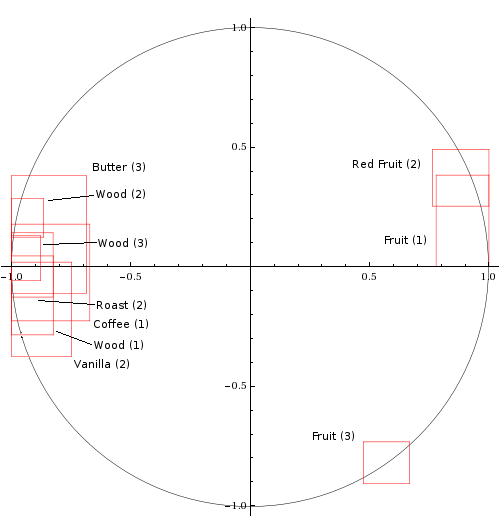}\\
(a) Correlations amongst tables & (b) Evolution of the variables\\
\includegraphics[width=3cm]{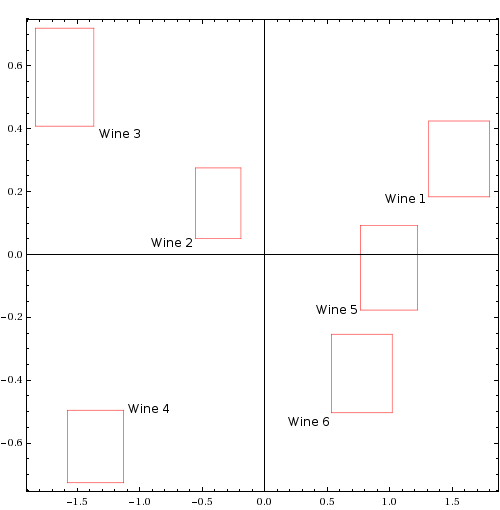} & \includegraphics[width=3cm]{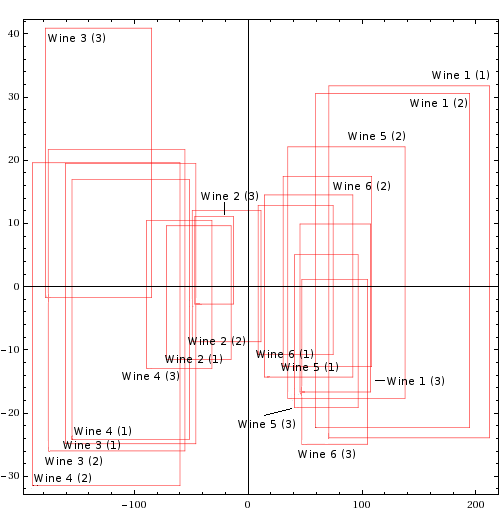}\\
(c) Average individuals & (d) Evolution of the individuals\\
\end{tabular}
\caption{Output of the \texttt{INTERSTATIS} method.}
\end{center}
\end{figure}

Analyzing figure 1a, the test data produces a very similar output
to that of the \texttt{STATIS}, given that experts 1 and 2 are correlated
given their similitude, leaving expert 3 on its own. Nonetheless,
the distance between the groups of tables is variable, as a consequence
of the use of interval data. For example, the closeness of the borders
for experts 2 and 3 indicates a similar behavior in said borders, nonexistent
in the classic case. It is precisely this capability of handling a level of
uncertainty, the advantage of this symbolic proposal, since it analyzes
not only the variation amongst individuals, but their internal variation
as well.

Figure 1b presents a very similar correlations structure in comparison to
the one generated by the \texttt{STATIS}, showing three groupings of
variables, but in the symbolic case overlapping occurs amongst said groups.
It should be noted that the representations for the variables in 1b
are near the borders of the circle, clear sign of marked correlations. If follows
that results drawn upon them are thus valid.

In the average individuals plane (figure 1c) the additional information
provided by the \texttt{INTERSTATIS} can be appreciated, through the size
of the rectangles, given that their size is proportionate to the average
size of the represented individual.

In the last step (figure 1d) the evolution of the individuals is studied,
that is, their changes throughout all the tables or measurings. Given the amount of
rectangles present ($r \times n$), this plane should ideally be plotted
using a subset of the individuals, to aid in its interpretation. For example,
visualize the evolution of a pair of individuals to better appreciate
their variation and interaction.

It is important to remark that the symbolic proposal adds analysis information
without sacrificing that already provided by the classic method, that is, produces
more information by the use of intervals. Nevertheless, the interpretation complexity
is increased as well. For this proposal, an aspect that heavily affects the output of the
method is the width of the input intervals and therefore, the input tables should be normalized
so that the width of its elements is reduced and the output remains useful.

Another relevant characteristic of symbolic analysis, which adds value to the proposal,
lies in its ability to reveal possible interactions between individuals or variables,
hidden in the results obtained through the classic methods.

\section{Conclusions and future work}
A new proposal for applying the \texttt{STATIS} method when the input data consists
of interval valued data was proposed. This proposal permits the method to handle internal variation in
the input data, therefore allowing the use of representations which better reflect
the true nature of that which is measured. It was also proved that the \texttt{STATIS}
is a particular case of the proposed \texttt{INTERSTATIS} with IA and CPCA, thus
validating its statistical usefulness.

Another important benefit resides in the use of symbolic objects \emph{per se},
given that they can be used to compact copious amounts of data, while
minimizing data loss during the process. Once these symbolic objects
are created, the \texttt{INTERSTATIS} --as well as any other symbolic method--
allows the exploration and analysis of the totality (or a majority) of the individuals
present in the original data table.

With regards to future work, although the obtained results have been positive,
there exist other interval arithmetics (and modifications of these) that could
help make the output of the \texttt{INTERSTATIS} more compact.
Finally, a generalization of the \texttt{STATIS} method
to treat histogram valued data is an open problem.

\end{document}